\documentclass[twocolumn,prd,10pt,showkeys,showpacs]{revtex4}  
\pdfoutput=1

\usepackage{graphicx}  
\usepackage{dcolumn}   
\usepackage{bm}        
\usepackage{amssymb}   
\usepackage{amsmath} 
\usepackage{enumerate}

\newcommand{\PRE}[1]{}

\newcommand{\ba}{\begin{align}}
\newcommand{\ea}{\end{align}}
\newcommand{\bann}{\begin{align*}}
\newcommand{\eann}{\end{align*}}
\newcommand{\bea}{\begin{eqnarray}}
\newcommand{\eea}{\end{eqnarray}}

\def\lsim{\mathrel {\vcenter {\baselineskip 0pt \kern 0pt \hbox{$<$} \kern 0pt \hbox{$\sim$} }}}

\begin{document}

\title{Sterile Neutrino Production Through a Matter Effect Enhancement
at Long Baselines}

\author{Joseph Bramante}

\affiliation{Department of Physics and Astronomy, University of
Hawaii, \\ 2505 Correa Rd., Honolulu HI, USA}

\email{bramante@hawaii.edu}

\begin{abstract}
If sterile neutrinos have a neutral coupling to standard model fermions, matter effect resonant transitions to sterile neutrinos and excess neutral-current events could manifest at long baseline experiments. Assuming a single sterile neutrino with a neutral coupling to fermionic matter, we re-examine bounds on sterile neutrino production at long baselines from the MINOS result $P_{\nu_{\mu} \rightarrow \nu_s} < 0.22$ (90\% CL). We demonstrate that sterile neutrinos with a neutral vector coupling to fermionic matter could evade the MINOS limit, allowing a higher fraction of active to sterile neutrino conversion at long baselines. Scanning the parameter space of sterile neutrino matter effect fits of the LSND and MiniBooNe data, we show that in the case of a vector singlet coupling of sterile neutrinos to matter, some favored parametrizations of these fits would create neutral-current event excesses above standard model predictions at long baseline experiments (e.g. MINOS and OPERA).
\end{abstract}
\date{\today}


\preprint{UH-511-1181-2011}


\maketitle

\section{Introduction}
Although many proposals of extra generations of neutrinos apply to small neutrino mixing anomalies, the first detection of a sterile neutrino could come in the form of a very large mixing anomaly at a long baseline. 

A recent result from the OPERA experiment \cite{opera} measured a difference in 
the speed of light in vacuum and the speed of muon neutrinos along a 730 km baseline:
$
(v_{\nu_{\mu}}-c)/c=(2.48 \pm 0.41)\times10^{-5}~~\rm{[OPERA]}
$.
This data turned out to be an experimental error. Nevertheless, the OPERA result prompted proposals of new physics\cite{newphys,kostelecky1,kostelecky2,kostelecky3,kostelecky4} and phenomenological constraints on muon neutrino Lorentz invariance violation \cite{newphys,Winter,Cohen}. One popular superluminal mechanism for neutrinos involved sterile neutrino transport through a higher dimensional bulk \cite{Nicolaidis:2011eq,Hannestad:2011bj,giu, Pas:2005rb,Dent:2007rk}, which supposed active neutrinos confined to a D3 brane oscillate to sterile neutrinos, whose lack of gauge charge leaves them free to travel through large extra dimensions. However, a strong constraint on these sterile neutrino models comes from measurements of neutrinos and photons arriving from SN1987a. The detection of 24 neutrino events at three sites \cite{sn1987a1,sn1987a2,sn1987a3} arriving $\sim$ 4 hours before SN1987a photons puts a rather stringent bound on superluminal electron neutrinos, $(v_{\nu_e}-c)/c~\sim~3\times10^{-9}~~\rm{[IMB,KII,Baksan]}.$ Although OPERA detected muon neutrinos and SN1987a produced electron neutrino data, leading to the possibility of a flavor anomaly, additional experimental constraints on neutrino mass eigenstate velocity differences \cite{giu,giuflavor1,giuflavor2,giuflavor3,giuflavor4}
ruled out active flavour-dependent velocity anomalies, short of replacing the standard PMNS matrix with a different formalism \cite{kostelecky1,kostelecky2,kostelecky3,kostelecky4}. 

Although it is settled that the OPERA experiment did not observe a superluminal anomaly \cite{opera}, nevertheless some of the phenomenological studies of sterile neutrino production at long baselines are applicable to future neutrino studies.
In particular, among the many OPERA constraints papers that arose in the wake of the anomalous OPERA data \cite{Winter} pointed out that the fraction of sterile neutrinos required to explain the anomaly was at odds with a prior study of sterile neutrinos at MINOS \cite{minos}. 

In this paper we show that if sterile neutrinos have non-standard interactions with fermionic matter, this induces matter effect resonance transitions and excess neutral-current events that would simultaneously allow for the MINOS result and a sizeable production of sterile neutrinos at another long baseline. We develop the phenomenology of a matter-dependent increase of sterile neutrino production through a sterile neutrino neutral U(1) vector coupling to fermions. Similar models employing new sterile neutrino interactions via a B-L gauge boson have been developed in \cite{NW1,NW2,NW3,NW4} to fit neutrino disappearance anomalies at short and long baseline experiments \cite{sb1,sb2,sb3,sb4} . In this paper we consider a sterile neutrino matter effect model for which sterile neutrino neutral interactions with matter would be detectable.

The structure of this paper is as follows: In section II we find the evolution equation and transition probability for a sterile neutrino with a new neutral U(1) coupling to fermions. In section III we examine constraints on sterile neutrinos with neutral current interactions and additionally comment on specific constraints on sterile neutrinos as an explanation for the OPERA anomaly. In section IV we determine what parametrizations of sterile neutrino mass and coupling to standard model fermions would produce detectable abundances at current long baseline studies. Additionally we compare these parametrizations with other studies fitting a similar model to short baseline anomalies. In section V we conclude. 

\section{Sterile Neutrinos with a Neutral Vector Singlet Coupling} \label{ii}

Sterile neutrino models usually suppose that the active neutrinos have a small mixing angle with extra generations of non-interacting, or sterile neutrinos. Here we examine how these models change with the introduction of a straightforward U(1) vector coupling of sterile neutrinos to standard model fermions. In short, we find that the coupling causes an enhancement of an otherwise small muon-sterile mixing term in matter.
\begin{figure} 
\includegraphics[scale=.42]{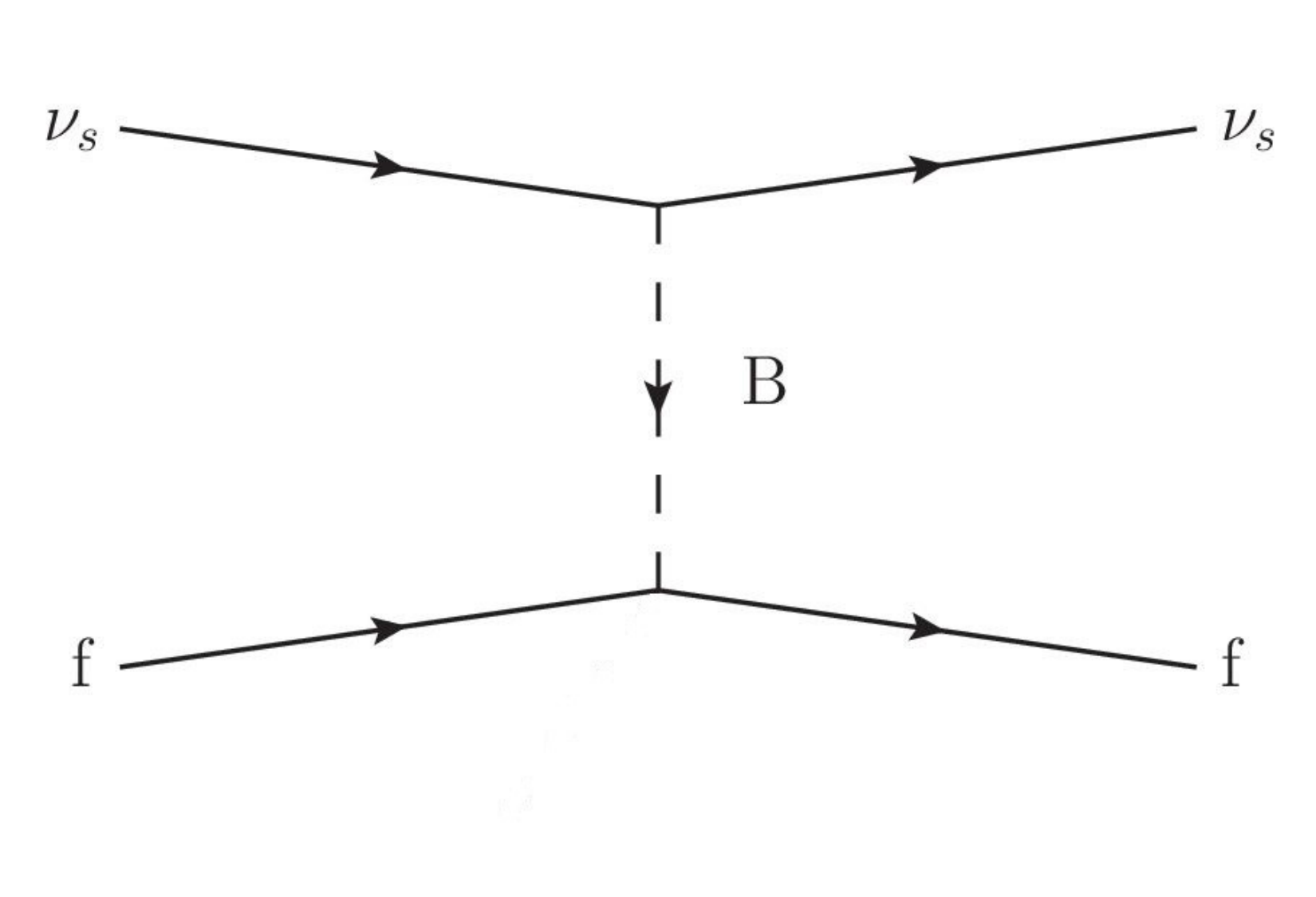}
\caption{$\nu_s$ elastic scattering 
through a vector singlet $B$.}
\label{bexchange}
\end{figure}
The contribution of the process in figure \ref{bexchange} to the effective potential of sterile neutrino propagation is given by
\begin{equation}
\label{hami}
\mathcal{H}_{eff}^{(B)}=- \frac{g_{s}g_f}{8m_B^2}[\bar{\nu_s} \gamma^{\mu} \nu_{s}]
[\bar{f_{R,L}} \gamma_{\mu} f_{R,L}],
\end{equation}
where $f$ is a fermion abundant in matter, e.g. $\rm{(e^-,u,d)}$, and B is a neutral vector boson singlet. In general, B will couple to $\rm{(u,d)_L,(\nu_{e},e)_L,u_R,d_R,~and~e_R}$ with strength $g_f$ and couple to $\rm{\nu_{s}}$ with strength $g_s$. Depending on the scale of \eqref{hami}, the couplings to sterile neutrinos and fermions will need to be unequal to avoid precision electroweak constraints on standard model fermion couplings to a new vector singlet boson ($g_f$).

The proposed active-sterile matter mixing enhancement will affect all active-sterile neutrino oscillations in matter, but to simplify our analysis and focus on parameters which maximize muon neutrino oscillation to sterile neutrinos, we here set the vacuum mixing angles for $\nu_{e,\tau} \leftrightarrow \nu_s$ to zero, so that the only appreciable sterile neutrino production in matter will come from $\nu_{\mu} \leftrightarrow \nu_s$. For this simple system with only $\nu_{\mu}$ mixing with $\nu_s$, the flavor evolution equation in matter is \cite{mswmech1,mswmech2,mswmech3,mswmech4}
\begin{eqnarray}
\label{flavorevolution}
&i& \frac{d}{dt} \left( \begin{array}{c} A_{\nu_{\mu} \rightarrow \nu_{\mu}} 
\\ A_{\nu_{\mu} \rightarrow \nu_{s}}  \end{array}\right)
= \nonumber \\
& &\left( \begin{array}{cc} 
-\frac{\Delta m^2}{4E}\cos2\theta - \sqrt{2}N G_s & \frac{\Delta m^2}{4E}\sin2\theta 
\\ \frac{\Delta m^2}{4E}\sin2\theta & \frac{\Delta m^2}{4E}\cos2\theta + \sqrt{2}N G_s
\end{array}\right)
\nonumber \\ &\times& \left( \begin{array}{c} A_{\nu_{\mu} \rightarrow \nu_{\mu}} 
\\ A_{\nu_{\mu} \rightarrow \nu_{s}}  \end{array}\right), 
\end{eqnarray}
where $G_s \equiv \frac{\sqrt{2}g_{s}g_f}{8m_B^2}$, $N=n_e+n_u+n_d$ is the number density of matter fermions, $\theta$ is the vacuum mixing angle and $\Delta m^2$ is the squared mass difference between the mass eigenstates of $\nu_{\mu}$ and $\nu_s$ in vacuum. Standard model MSW terms in (\ref{flavorevolution}) have a small effect over a $\sim 10^3$ km baseline and have been omitted. Diagonalizing the evolution Hamiltonian yields
\begin{eqnarray}
\frac{\Delta m_M^2}{4E} \left( \begin{array}{cc} 
-\cos2\theta_M & \sin2\theta_M
\\ \sin2\theta_M & \cos2\theta_M
\end{array}\right)
\end{eqnarray}
where the mixing angle and squared mass difference in matter are
\begin{eqnarray}
\sin 2\theta_M &=& \frac{\Delta m^2}{\Delta m_M^2}\sin 2\theta
\\
\Delta m_M^2 &=& \Delta m^2 \left( \left(\cos 2\theta - 
\frac{2\sqrt{2}NE G_s}{\Delta m^2}\right)^2 +\sin ^2 2\theta \right)^{1/2} \nonumber \\
\end{eqnarray}
and the corresponding matter $\nu_{\mu} \rightarrow \nu_s$ transition 
probability over a distance $D$ is
\begin{eqnarray}
P_{\nu_{\mu} \rightarrow \nu_s} &=& \sin^2 2\theta_M \sin^2 
\left(\frac{\Delta m_M^2 D}{4E} \right) \nonumber
\\ &=& \frac{\sin^2 2\theta}
{\left(\cos 2\theta - 
\frac{2\sqrt{2}NE G_s}{\Delta m^2}\right)^2 +\sin ^2 2\theta}
\nonumber \\ &\times &\sin^2\left( \frac{\Delta m^2  D}{4E} \sqrt{ \left(\cos 2\theta - 
\frac{2\sqrt{2}NE G_s}{\Delta m^2}\right)^2 +\sin ^2 2\theta } \right).\nonumber
\\
\label{prob}
\end{eqnarray}

\section{Constraints on Sterile Neutrinos With a Vector Singlet Coupling}

A strong bound on muon neutrino oscillation to sterile neutrinos in matter comes from the MINOS measurement of neutral-current (NC) interactions of the NuMI muon neutrino beam at the end of a 730 km baseline \cite{minos}. The MINOS result of 802 NC events against an expected $754 \pm 28_{\rm{stat}} \pm 37_{\rm{sys}}$ event background excludes $P_{\nu_{\mu} \rightarrow \nu_s} > 0.22$ at 90\%\ confidence. However, in the case of the sterile neutrino matter effect model considered here, there is an additional contribution to NC events from the coupling of $\nu_s$ to standard model fermions (\ref{hami}). If the scale of the interaction considered is on the order of the Fermi constant, $G_s \approx G_F$, sterile neutrino interactions with standard model fermions would contribute to the neutral current event counts at long baseline experiments. 

Assuming that the mass of the new neutral vector boson is much greater than the momentum of the sterile neutrino, at current long baseline energies the four-fermi approximation is valid for active, $\sigma_{NC\nu_{a}}\propto G_F E_{\nu}^2 N_e$, as well as sterile neutrino neutral current interactions, $\sigma_{NC\nu_{s}} \propto 7 G_s E_{\nu}^2 N_e$, where $N_e$ is the density of electrons in matter. The contribution of a sterile neutrino to the neutral current event rate will be equal to the contribution of an active neutrino multiplied by a proportionality constant $\alpha \equiv 7 G_s/G_F$,  where the factor of 7 arises because the singlet vector coupling of $\nu_s$ to matter does not have V-A diagram cancellations \cite{mswmech1,mswmech2,mswmech3,mswmech4}. Thus one way to construct a sterile neutrino matter effect model that allows for $P_{\nu_{\mu} \rightarrow \nu_s} \gtrsim 0.30$ and is consistent with the MINOS measurement is to set $G_s \gtrsim G_F/7$ and $m_B >> E_{\nu}$.

While this study uses a single active-sterile mixing angle and squared mass difference to identify possible active to sterile mixing resonances at long baselines, any modification of muon neutrino mixing in matter is subject to constraints from measurements of the atmospheric mixing angle \cite{Wendell:2010md}. Most parametrizations of this model are ruled out by these measurements. However, very small vacuum mixing angles would create active to sterile mixing resonances over a small range of neutrino energies, as shown in figure \ref{PE}. With a small enough vacuum mixing angle, it would be possible to identify a resonance at a baseline experiment, while the signal of this neutrino disappearance (and extra flux of neutral current events) would not be evident in more broadly binned energy data at atmospheric experiments.

\subsection{Constraints on Superluminal Sterile Neutrinos at OPERA}

Although the OPERA anomaly was an experimental error, nevertheless there is continued interest in Lorentz-violating neutrinos \cite{kostelecky1,kostelecky2,kostelecky3,kostelecky4} and the constraints developed for OPERA may be applicable to future neutrino anomalies. Here we briefly evaluate constraints on the OPERA superluminal anomaly for sterile neutrinos with a neutral vector coupling. 

In \cite{Winter} it was demonstrated that there is a minimum fraction of neutrinos which must travel superluminally in order to reproduce the OPERA anomaly. The spectral flatness of time-binned neutrino events requires the superluminal fraction $ \chi = \Sigma_{\nu_{c+}}|U_{\mu \rightarrow \nu_{c+}}|^4 / \Sigma_i |U_{\mu i}|^4$ to be at least $\sim$ 0.18 at $3 \sigma$ and 0.28  at $2 \sigma$ confidence. Furthermore, \cite{Cohen} showed that superluminal \emph{active} neutrinos would undergo $\nu_f \rightarrow \nu_f ~e^+ ~e^-$ Cherenkov-like radiation forcing an effective energy cutoff above $\sim$ 12.5 GeV for active neutrinos travelling 730 km at 7.5 km/s faster than light. To avoid this energy cutoff, we might decide to require that all superluminal propagation occur through sterile neutrinos. This requirement combined with the neutrino fraction constraint demands $P_{\nu_{\mu} \rightarrow \nu_s} > 0.18$, though a more promising model would allow for $P_{\nu_{\mu} \rightarrow \nu_s} \gtrsim 0.30$. Although we previously specified that the mixing $\nu_{e} \rightarrow \nu_s$ would be very small, in principle there must be some mixing of all active neutrinos with sterile through shared mass eigenstates. Thus even a weak scale coupling ($G_s \approx G_F$) reintroduces the Cerenkov radiation cutoff problem, because the superluminal sterile neutrino will oscillate to electron flavor.

\begin{figure}
\includegraphics[scale=1]{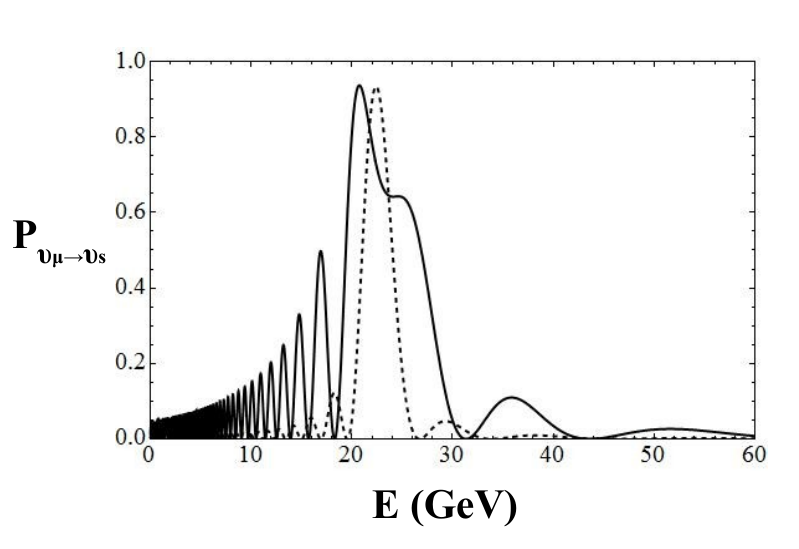}
\caption{Sterile neutrino transition probability plotted against neutrino energy for parameters indicated in the text.}
\label{PE}
\end{figure}

The OPERA experiment completed an additional study in which the proton bunches and resulting neutrino packets were more tightly spaced in time \cite{opera}. The analysis in \cite{Winter} was updated to show that
\begin{eqnarray}
P_{\nu_{\mu} \rightarrow \nu_s} &\gtrsim & 0.40~~~6\sigma ~\rm{exclusion} \nonumber \\
P_{\nu_{\mu} \rightarrow \nu_s} &\gtrsim & 0.80~~~3\sigma ~\rm{exclusion}.
\end{eqnarray}
Although a sterile neutrino fraction of this magnitude may seem unrealistic, one interesting feature of sterile neutrinos with a weak scale vector singlet interaction is that at terrestrial baseline lengths and energies, there are parametrizations which cause energy-dependent resonant transitions $P_{\nu_{\mu} \rightarrow \nu_s} \sim 1$. In figure \ref{PE} we plot the transition probability \eqref{prob} for the parameters $\Delta m^2 = 0.45 ~{\rm eV^2},~A_s=10^{-11} ~{\rm eV},~\sin^2 2\theta=0.05 {\rm ~(solid)},~\sin^2 2\theta=0.005 ~{\rm (dashed)},$ and $~D=730~{\rm km}$. If sterile neutrinos have a neutral-current interaction cross-section seven times larger than active neutrino NC interactions, an oscillation of transition probability peaking at 5\%\ in figure \ref{PE} would be observable as an oscillation of NC event flux of $\sim \pm 20 \%$ around SM expectations.

\section{Parametrization for Resonance at Long Baselines}

Sterile neutrino matter effects have recently been considered as an explanation of short and long baseline anomalies \cite{NW1,NW2,NW3,NW4,kara1,kara2}. Most fits indicate a 4th neutrino with a mass of about 0.5 eV. Particularly, \cite{kara1,kara2} uses a ``model agnostic" $\nu_s$ matter effect to fit a 3+1 model to the LSND and MiniBooNE datasets. The active-sterile mixing angles, matter effect potentials, and squared mass differences of \cite{kara1,kara2} are consistent with a parametrization which would cause substantial $\nu_{\mu} \rightarrow \nu_s$ mixing at long baseline experiments for the model we outline here. 

To demonstrate what parameters are required to create a large active to sterile matter effect resonance for the simplified mixing equations here, we will use the OPERA baseline and neutrino energies and require $P_{\nu_{\mu} \rightarrow \nu_s} \gtrsim 0.30$ for some set of neutrino energies. To avoid obviously contradicting muon neutrino data at non-resonant energy ranges, we set the vacuum mixing angle $\sin^2 2\theta=0.05$, which over very long distances implies a vacuum $\nu_{\mu} \rightarrow \nu_s$ transition probability of 0.025. Inspection of the first term in $P_{\nu_{\mu} \rightarrow \nu_s}$(\ref{prob})
\begin{equation}
\frac{\sin^2 2\theta}
{\left(\cos 2\theta - 
\frac{2\sqrt{2}NE G_s}{\Delta m^2}\right)^2 +\sin ^2 2\theta}
\end{equation}
produces a squared-mass difference-coupling resonance condition
\begin{equation}
\frac{2\sqrt{2}NE G_s}{\Delta m^2} = 
\frac{2EA_s}{\Delta m^2} \sim \mathcal{O} (1);
\label{con1}
\end{equation}
If $\frac{2EA_s}{\Delta m^2} >> 1$, $P_{\nu_{\mu} \rightarrow \nu_s}$ will diminish rapidly, and if $\frac{2EA_s}{\Delta m^2}<<1$, $P_{\nu_{\mu} \rightarrow \nu_s}$ cannot exceed a value of 0.05. Inserting the OPERA values into the second term of $P_{\nu_{\mu} \rightarrow \nu_s}$, D = 730 km and E $\sim$ 17 GeV,
\begin{equation}
\sin^2\left(\Delta m^2 54 eV^{-2} \sqrt{\left(\cos 2\theta - 
\frac{2\sqrt{2}NE G_s}{\Delta m^2}\right)^2 +\sin ^2 2\theta } \right)
\label{con2}
\end{equation}
a minimum value of $\Delta m^2$ (0.04 $\rm{eV}^2$) becomes apparent. This bound follows from the maximum value of the prefactor of \eqref{prob}, which is unity. If
\begin{eqnarray}
\left(\cos 2\theta - 
\frac{2\sqrt{2}NE G_s}{\Delta m^2}\right)^2 \sim 0 \nonumber
\end{eqnarray}
then the prefactor equals
\begin{eqnarray}
\frac{\sin^2 2\theta}{\left(\cos 2\theta - 
\frac{2\sqrt{2}NE G_s}{\Delta m^2}\right)^2 +\sin ^2 2\theta} \sim 1 \nonumber
\end{eqnarray}
leading to a maximum transition probability of
\begin{eqnarray}
P_{\nu_{\mu} \rightarrow \nu_s} =\sin^2\left(2.2 \sqrt{0.05} \right) \sim 0.30.
\end{eqnarray}
As $\Delta m^2$ increases substantially from this value, the second term in $P_{\nu_{\mu} \rightarrow \nu_s}$ will average to $\frac{1}{2}$ and the first term will have to resonate at $\sim \frac{2}{3}$ to produce $P_{\nu_{\mu} \rightarrow \nu_s}$ = 0.30.

\begin{figure}
\includegraphics[scale=.75]{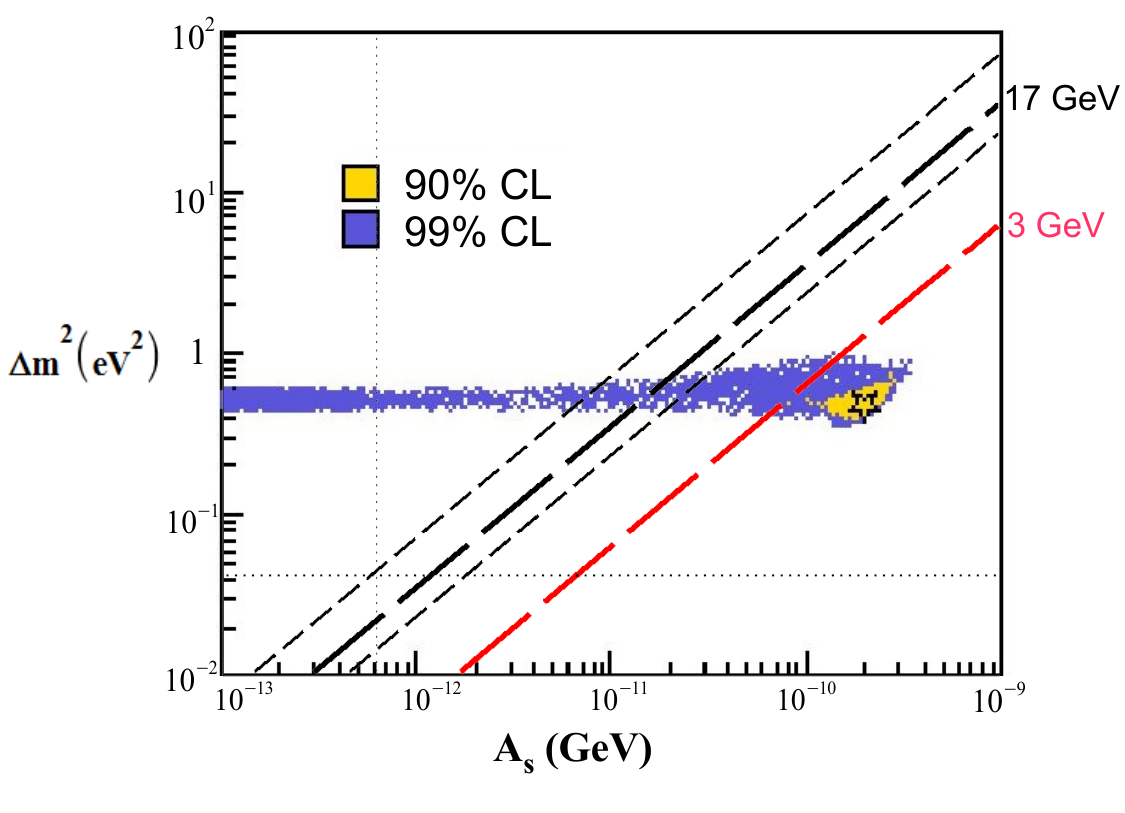}
\caption{Lines of maximum resonant active to sterile mixing for sterile neutrino matter effect models as detailed in the text. The underlaid scatter plot fit is taken from \protect\cite{kara1,kara2}.}
\label{fits}
\end{figure}

In figure \ref{fits} have plotted a band of black dashes which includes the region $P_{\nu_{\mu} \rightarrow \nu_s} > 0.18$ to illustrate the resonance $\Delta m^2 \sim 2EA_s$. The band of black dashes centers on the maximum resonance $(0.975)\Delta m^2 = 2E_{\nu}A_s$ for 17 GeV muon neutrinos, for which the transition probability $P_{\nu_{\mu} \rightarrow \nu_s}$ can approach unity. The same curve is plotted for 3 GeV neutrinos, which is the central neutrino energy at MINOS. The dotted vertical line indicates the value of $A_s$ which corresponds to sterile neutrinos having weak scale neutral interactions with standard model fermions. The dotted horizontal line indicates the smallest possible $\Delta m^2$ value which yields $P_{\nu_{\mu} \rightarrow \nu_s} \gtrsim 0.30$ for 17 GeV neutrinos when $\sin^2 2\theta=0.05$. Underlaid is a scatter plot taken from \cite{kara1,kara2}, which fits $\Delta m_{41}^2$ and $A_s = \sqrt{2}G_sN$ to LSND and MiniBooNE neutrino oscillation data. It should also be noted when examining figure \ref{fits} that the choice of $\sin^2 2\theta$ affects the location and shape of the resonance lines. Increasing the vacuum sterile mixing angle from the value $\sin^2 2\theta = 0.05$ will both elevate and broaden the $P_{\nu_{\mu} \rightarrow \nu_s} > 0.18$ inclusion band.

If sterile neutrinos interact via neutral-currents with all matter fermions, in the case of a straightforward singlet vector coupling the parameter space around the 3 GeV (MINOS) resonance line is certainly ruled out. (MINOS was not bombarded with NC events in \cite{minos}). However, there are intermediate values of $A_s$ along and beside the 3 GeV line, where the cross-section for sterile neutrino NC events exceeds that of active neutrinos, and sterile neutrinos are produced in significant quantities. A possible signal of this in a long baseline neutral current event study would be an oscillation above the SM background of NC event counts with respect to energy (see figure \ref{PE}).

\section{Conclusion}

In conclusion we have developed the phenomenology of a matter effect enhanced model of interacting sterile neutrinos at long baselines. We have shown that for parameters commonly used in sterile neutrino matter effect models of short and long baseline anomalies, sterile models with a neutral vector singlet coupling could be observed at a long baseline experiment as a severe over-production of neutral current events or as an oscillation of neutral current events over the expected SM background. We demonstrated that this model avoids prior constraints \cite{Winter} on sterile neutrinos as an explanation for the now defunct OPERA superluminal anomaly.

While we have shown what signals might arise in the case of a sterile neutrino coupling to standard model fermions via a neutral vector singlet boson, a more complete treatment of this model would need to demonstrate that the new coupling arises from a fully renormalizable theory \cite{koppetal1,koppetal2,koppetal3}. In addition, it is not clear whether the model outlined here would be consistent with atmospheric neutrino data under any parametrization. We leave the fitting of neutral-coupled sterile neutrinos to atmospheric, baseline, and reactor neutrino data to future work.

\begin{acknowledgments}
We wish to thank Jason Kumar, John Learned, Danny Marfatia, Sandip Pakvasa, Heinrich Pas,
and David Yaylali for useful discussions. This work is supported in part by 
by Department of Energy grant DE-FG02-04ER41291.
\end{acknowledgments}

\end{document}